\documentclass[aps, prd,  showpacs ]{revtex4}
\usepackage{amsmath,amsfonts,amscd,amssymb,epsf,color}

\newcommand{\vep}{\varepsilon}
\newcommand{\pa}{\partial}
\newcommand{\be}{\begin{equation}}\newcommand{\ee}{\end{equation}}
\newcommand{\Ref}[1]{(\ref{#1})}

\begin{document}

\title{On the corrections beyond proximity force approximation (PFA)}
\author{L. P. Teo}\email{LeePeng.Teo@nottingham.edu.my}
    \affiliation{Department of Applied Mathematics, Faculty of Engineering, University of Nottingham Malaysia Campus, Jalan Broga, 43500, Semenyih, Selangor Darul Ehsan, Malysia}
\author{M.~Bordag\email{bordag@itp.uni-leipzig.de}}
    \affiliation{Leipzig University, Vor dem Hospitaltore 1, D-04103 Leipzig, Germany}
\author{V.~Nikolaev\email{Vladimir.Nikolaev@ide.hh.se}}
    \affiliation{Halmstad University, Box 823, 30118 Halmstad, Sweden}

\begin{abstract}
We recalculate the first analytic correction beyond PFA for a sphere in front of a plane for a scalar field and for the electromagnetic field. We use the method of Bordag and Nikolaev [J.Phys.A, {\bf 41} (2008) p.164002]. We confirm their result for Dirichlet boundary conditions whereas we find a different one for Robin, Neumann and conductor boundary conditions. The difference can be traced back to a sign error. As a result, the corrections depend on the Robin parameter. Agreement is found with a very recent method of derivative expansion.
\end{abstract}
\maketitle

\section{Introduction}
The proximity force approximation (PFA) is one of the basic methods for calculation of Casimir and van der Waals forces between non-planar surfaces. Although it appeared more than 70 years ago, corrections beyond PFA, which gives the leading order only, were not known. Only with the new method of the Casimir force calculation,  based on the multiple scattering formalism in conjunction with Krein's formula \cite{bulg06-73-025007} or on a path integral quantization with a partial wave expansion \cite{emig06-96-080403}, it became possible to calculate corrections beyond PFA. Using an asymptotic expansion, analytic corrections were calculated, first for a cylinder in front of a plane \cite{bord06-73-125018}, later for a sphere in front of a plane \cite{bord08-41-164002,bord10-81-065011}. At the same time, numerical approximations to these corrections were obtained \cite{gies06-74-045002}.

The method, used in \cite{bulg06-73-025007,emig06-96-080403}, results in a representation of the vacuum energy for a sphere in front of a plane (see Fig. \ref{f1}),
\be\label{E0}E=\frac{1}{2\pi}\int_0^\infty d\xi\, \text{Tr}\ln\left(1-\mathbf{M}\right),
\ee
where $\xi$ is the imaginary frequency  and $\mathbf{M}$ is composed of the scattering T-matrix of the sphere (see below). In \Ref{E0}, the trace is over  orbital momenta, $l=0,1,\dots$ and $m=-l,\dots,l$. In case the ratio
\be\label{ep}\vep=\frac{d}{R}
\ee
of the separation to the radius of the sphere is not small, the sums and the integral in \Ref{E0} converge rapidly allowing for an easy numerical evaluation of the vacuum energy and of the Casimir force. On the contrary, if $\vep$ becomes small, i.e. for close separation, higher orders of the orbital momenta need to be accounted for. As a consequence, the numerical evaluation could be done down to $\vep\sim 0.1$ only.

The analytic method \cite{bord06-73-125018,bord08-41-164002,bord10-81-065011} makes an asymptotic expansion of \Ref{E0} for $\vep\to0$. This involves a substitution of the sums by integrals, a change of variables and an asymptotic expansion of $\mathbf{M}$ (involving Bessel functions and Clebsch-Gordan coefficients).

For a cylinder in front of a plane, this method was verified independently in  \cite{teol11-84-025022} (and generalized to finite temperature). Regrettably, for a sphere in front of a plane, for Neumann boundary conditions and for the electromagnetic case,  the results of \cite{bord08-41-164002,bord10-81-065011} could not be confirmed. In fact,  a sign error was found and it is the aim of the present paper, to point to the place where it occurred and to give the correct results following from this method.

% However, for a sphere in front of a plane, for Neumann boundary conditions and for the electromagnetic case, an error in the results of \cite{bord08-41-164002,bord10-81-065011} was found. It is the aim of the present paper, to point this out and to give the correct results following from this method.
%

In the following two sections the corrections beyond PFA are recalculated for the scalar field and for the electromagnetic field. The last section contains the conclusions together with a comparison with further methods.

\noindent Throughout the paper we use units with $\hbar=c=1$.
%%
%\be\label{}
%\ee
%%
%%
%\be\label{}
%\ee
%%

\section{Scalar field}
\begin{figure}[h]
\epsfxsize=0.5\linewidth \epsffile{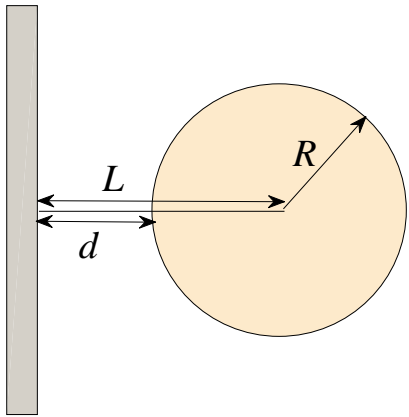} \caption{\label{f1} The configuration of a sphere in front of a plane. }\end{figure}
Following \cite{bord10-81-065011}, let the  radius of the sphere be $R$, and   the distance from the center of the sphere to the plane be $L$. Then the distance between the sphere and the plane is $L-R$, which we denote   by $d$ (see Fig. \ref{f1}).

For a scalar field $\varphi$, the Casimir interaction energy is given by \cite{bord08-41-164002,bord10-81-065011,emig06-96-080403}:
\begin{equation}\label{eq10_13_1}
E^{\text{XY}}=\frac{1}{2\pi}\int_0^{\infty}d\xi\,\text{Tr}\ln \left(1-(-1)^xN^{\text{Y}}(\xi)\right).
\end{equation}
Here we follow the notations in \cite{bord10-81-065011}. The first index denotes the boundary conditions on the plane. For Dirichlet boundary conditions, $\text{X}=\text{D}$ and $x=0$. For Neumann boundary conditions, $\text{X}=\text{N}$ and $x=1$. The second index denotes the boundary conditions on the sphere.  $\text{Y}=\text{D}, \text{N}$ and  $\text{R}$ for Dirichlet, Neumann and general Robin boundary conditions respectively. The trace $\text{Tr}$ is the orbital momentum sum,
$$\text{Tr}= \sum_{m=-\infty}^{\infty}\sum_{l=|m|}^{\infty}, $$
and the matrix $N^{\text{Y}}$  is given by
\begin{equation}\label{eq10_13_21}
N_{l,l'}^{\text{Y}}(\xi)=\sqrt{\frac{\pi}{4\xi L}}\sum_{l^{\prime\prime}=|l-l'|}^{l+l'}K_{l^{\prime\prime}+1/2}(2\xi L)
H_{ll' }^{l^{\prime\prime}}d_l^\text{Y}(\xi R).
\end{equation}
Here,
\begin{equation}\label{eq10_13_20}
H_{ll'}^{l^{\prime\prime}}=\sqrt{(2l+1)(2l'+1)}(2l^{\prime\prime}+1)\begin{pmatrix} l & l' & l^{\prime\prime}\\ 0 & 0 & 0\end{pmatrix}
\begin{pmatrix} l & l' & l^{\prime\prime}\\ m & -m & 0\end{pmatrix},
\end{equation}involves $3j$-symbols and the function $d_l^\text{Y}(\xi R)$ depends on the boundary conditions on the sphere. For Dirichlet boundary conditions, i.e., $\varphi\bigr|_{r=R}=0$, it is given by
\begin{equation*}
d_l^{\text{D}}(\xi R)=\frac{I_{l+1/2}( \xi R)}{K_{l+1/2}( \xi R)},
\end{equation*}
whereas for general Robin boundary conditions with parameter $\alpha$, i.e., $r\pa_r\varphi+\alpha \varphi\bigr|_{r=R}=0$, it is given by
\begin{equation*}
d_l^{\text{R}}(\xi R)= \frac{uI_{l+1/2}( \xi R)+ \xi R I_{l+1/2}'( \xi R)}{uK_{l+1/2}( \xi R)+ \xi R K_{l+1/2}'( \xi R)}.
\end{equation*}
The parameter  $u$ is related to the Robin parameter $\alpha$ by $u=\alpha-1/2$. $\alpha=0$ or equivalently, $u=-1/2$, corresponds to the Neumann boundary conditions. In these formulas, $I_{\nu}(z)$ and $K_{\nu}(z)$ are the modified Bessel functions.

The asymptotic expansion for the Casimir energy when $\vep=d/R\ll 1$ can be computed in the same way as explained in \cite{bord08-41-164002,bord10-81-065011}. First make a substitution $\xi\mapsto \xi/R$ and expand the logarithm in \eqref{eq10_13_1} to obtain
\begin{equation}\label{eq10_13_2}
E^{\text{XY}}=-\frac{1}{2\pi R}\sum_{s=0}^{\infty}\frac{(-1)^{x(s+1)}}{s+1}\int_0^{\infty}d\xi \sum_{m=-\infty}^{\infty}\sum_{l=|m|}^{\infty}
\left(\prod_{j=1}^s \sum_{l_j=|m|}^{\infty}\right)\left(\prod_{i=0}^s N_{l_i,l_{i+1}}^{\text{Y}}\right).
\end{equation}
 The main contribution to the energy comes  from $l\sim l_1\sim \ldots \sim l_s$. Replacing $l_i, i=1,\ldots,s$ by $l+\tilde{l}_i$,   one can rewrite \eqref{eq10_13_2} as
\begin{equation}\label{eq10_13_3}
E^{\text{XY}}=-\frac{1}{2\pi R}\sum_{s=0}^{\infty}\frac{(-1)^{x(s+1)}}{s+1}\int_0^{\infty} d\xi\sum_{l=0}^{\infty}\sum_{m=-l}^{l}
\left(\prod_{j=1}^s \sum_{\tilde{l}_j=|m|-l}^{\infty}\right)\left(\prod_{i=0}^s N_{l+\tilde{l}_i,l+\tilde{l}_{i+1}}^{\text{Y}}\right),
\end{equation}
with the understanding that $\tilde{l}_0=\tilde{l}_{s+1}=0$. Using the Debye asymptotic expansions for the modified Bessel functions, one can show that
the leading contribution comes from $l\sim \vep^{-1}$, $\tilde{l}_i\sim \vep^{-\frac{1}{2}}$, $m\sim \vep^{-\frac{1}{2}}$ and $\xi\sim \vep^{-1}$.
This motivates to replace the summations by corresponding integrations, dropping only exponentially small contributions, and to make the substitutions,
\begin{equation}\label{eq10_14_2}\begin{split}
\xi=\frac{t}{\vep}\sqrt{1-\tau^2},&\hspace{1cm}l=\frac{t\tau}{\vep},\hspace{1cm}m=\sqrt{\frac{t\tau}{\vep}}\mu,\\
\tilde{l}_i=&\sqrt{\frac{4t}{\vep}}n_i\quad (i=1,\ldots, s).
\end{split}\end{equation}
One then obtains the following expression for the Casimir energy,
 \begin{equation}\label{eq10_13_4}
E^{\text{XY}}=-\frac{R}{4\pi d^2}\sum_{s=0}^{\infty}\frac{(-1)^{x(s+1)}}{s+1}\int_0^{\infty} dt \, t \int_0^1\frac{d\tau\sqrt{\tau}}{\sqrt{1-\tau^2}}\int_{-\sqrt{\frac{t\tau}{\vep}}}^{\sqrt{\frac{t\tau}{\vep}}}\frac{d\mu}{\sqrt{\pi}}
\left(\prod_{j=1}^s\int_{n_0}^{\infty}\frac{dn_j}{\sqrt{\pi}}\right)\mathcal{Z}^{\text{Y}},
\end{equation}
where
$$n_0=-\frac{\tau}{2}\sqrt{\frac{t}{\vep}}+\frac{1}{2}|\mu|\sqrt{\tau},$$and
\begin{equation}\label{eq10_13_6}\mathcal{Z}^{\text{Y}}=\prod_{i=0}^s\left(\sqrt{\frac{4\pi t}{\vep}}N^{\text{Y}}_{l+\tilde{l_i},l+\tilde{l}_{i+1}}\right).\end{equation}For the first two leading terms of the Casimir energy, one can set $\vep\rightarrow 0^+$ directly in the integration limits for $\mu$ and $n_j$, and expand $\mathcal{Z}$ up to terms of order $\vep$. For the term $H_{ll'}^{l^{\prime\prime}}$ defined  by \eqref{eq10_13_20}, it is nonzero only if $l+l'+l^{\prime\prime}$ is even. Using the substitution $$l^{\prime\prime}=l+l'-2\nu,$$ the summation over $l^{\prime\prime}$ in \eqref{eq10_13_21} becomes summation over $\nu$ from $\nu=0$ to $\nu=\min\{l,l'\}$. Since $l,l'\sim \vep^{-1}$, the upper limit for the $\nu$-summation can be replaced by $\infty$. For the $3j$-symbols, the small $\vep$ asymptotic expansion has been derived in \cite{bord08-41-164002}. For the modified Bessel functions, the small $\vep$ asymptotic expansion can be obtained using the Debye asymptotic expansions.  In the case the sphere is imposed with Dirichlet boundary conditions, i.e., $\text{Y}=\text{D}$,   the small $\vep$ asymptotic expansion is given by
\begin{equation*}
N_{l,l'}^{\text{D}}\sim \sqrt{\frac{\vep\tau}{2\pi t (1+\tau)}}e^{-2t-(n-n')^2}\int_{-\infty}^{\infty}\frac{d\eta}{\sqrt{\pi}}e^{-\eta^2+2i\eta\sqrt{2}\mu+\mu^2}\sum_{\nu=0}^{\infty}\frac{\eta^{2\nu}}{\nu!}
\left(\frac{1-\tau}{1+\tau}\right)^{\nu}\left(1+\sqrt{\vep}f_{n,n'}^{\text{D}}(\nu,\eta)+\vep g_{n,n'}^{\text{D}}(\nu,\eta)+\ldots\right).
\end{equation*}
One can then sum over $\nu$ and compute the Gaussian integration over $\eta$, which give
\begin{equation}\label{eq10_13_5}
N_{l,l'}^{\text{D}}\sim\sqrt{\frac{\vep}{4\pi t}}e^{-2t-(n-n')^2-\mu^2/\tau}\left(1+a_{n,n'}^{(1/2),\text{D}}\sqrt{\vep}+a_{n,n'}^{(1),\text{D}}\vep+\ldots\right).
\end{equation}The functions $a_{n,n'}^{(1/2),\text{D}}$ and $a_{n,n'}^{(1),\text{D}}$ are given in \cite{bord08-41-164002}.
Substituting \eqref{eq10_13_5} into \eqref{eq10_13_6}, one can expand  $\mathcal{Z}$ up to order $\vep$, which, upon substitution  into \eqref{eq10_13_4}  gives
 \begin{equation}\label{eq10_13_10}\begin{split}
E^{\text{XD}}\sim & -\frac{R}{4\pi d^2}\sum_{s=0}^{\infty}\frac{(-1)^{x(s+1)}}{s+1}\int_0^{\infty}dt \, t\,e^{-2t(s+1)} \int_0^1\frac{d\tau\sqrt{\tau}}{\sqrt{1-\tau^2}}\int_{-\infty}^{\infty}\frac{d\mu}{\sqrt{\pi}}\,e^{-\mu^2(s+1)/\tau}
\left(\prod_{j=1}^s\int_{-\infty}^{\infty}\frac{dn_j}{\sqrt{\pi}}\right)\\&
\hspace{3cm}\times e^{-\eta_1}\left(1+\left[\sum_{i=0}^sa_{n_i,n_{i+1}}^{(1/2),\text{D}}\right]\sqrt{\vep}+a^{\text{D}}\vep+\ldots\right),
\end{split}\end{equation}where
$\eta_1=\sum_{i=0}^s(n_i-n_{i+1})^2,$ and
\begin{equation}\label{eq10_14_4}
a^{\text{D}}=\sum_{0\leq i<j\leq s}a_{n_i,n_{i+1}}^{(1/2),\text{D}}a_{n_j,n_{j+1}}^{(1/2),\text{D}}+\sum_{i=0}^sa_{n_i,n_{i+1}}^{(1),\text{D}}.
\end{equation}
The integrations over $n_i$, $i=1,\ldots, s,$ can be performed in the same way as explained in \cite{bord06-73-125018}, with the help of a machine. The term proportional to $\sqrt{\vep}$ drops out since it is odd in one of the $n_i$'s. The integrations over $\mu, \tau$ and $t$ are   straightforward. In \cite{bord08-41-164002}, the following result was obtained for the case where Dirichlet conditions are imposed on the plane and the sphere ($\text{XY}=\text{DD}$):
\begin{equation}\label{eq10_13_7}
E^{\text{DD}}=
 -\frac{R}{16\pi d^2  }\sum_{s=0}^{\infty}\frac{1}{(s+1)^4}\left(1+\frac{\vep}{3}+\ldots\right)=-\frac{\pi^3 R}{1440 d^2}\left(1+\frac{\vep}{3}+\ldots\right).\end{equation}
Here the known formula
 $$\sum_{s=0}^{\infty}\frac{1}{(s+1)^4}=\zeta(4)=\frac{\pi^4}{90}$$ has been used.
 For the case where the plane is imposed with Neumann conditions and the sphere is imposed with Dirichlet conditions ($\text{XY}=\text{ND}$), we observe from \eqref{eq10_13_10} that the only difference is that now the summation over $s$ is alternating in sign. Hence we obtain immediately
 \begin{equation}\label{eq10_13_8}
E^{\text{ND}}=
 -\frac{R}{16\pi d^2  }\sum_{s=0}^{\infty}\frac{(-1)^{s+1}}{(s+1)^4}\left(1+\frac{\vep}{3}+\ldots\right)=\frac{7\pi^3 R}{11520 d^2  }\left(1+\frac{\vep}{3}+\ldots\right),\end{equation}
 where now the formula
  $$ \sum_{s=0}^{\infty}\frac{(-1)^{s+1}}{(s+1)^4}=-\frac{7}{8}\frac{\pi^4}{90}$$ is used.
The results \eqref{eq10_13_7} and \eqref{eq10_13_8} have been obtained in  \cite{bord08-41-164002} and \cite{bord10-81-065011}.

Next we consider the case where the sphere is imposed with general Robin boundary conditions. Following \cite{bord08-41-164002}, we observe that  the Debye asymptotic expansions give
\begin{equation}\label{eq10_13_12}\begin{split}
 uI_{\nu}(\nu z)+\nu z I_{\nu}'(\nu z)=&\nu \sqrt{1+z^2}  I_{\nu}(\nu z) \left(\frac{u}{\nu\sqrt{1+z^2}}+\frac{1+\frac{v_1}{\nu}+O\left(\frac{1}{\nu^2}\right)}{1+\frac{u_1}{\nu}+O\left(\frac{1}{\nu^2}\right)}\right)\\
 =&\nu \sqrt{1+z^2}  I_{\nu}(\nu z)\left(1+\frac{v_1-u_1}{\nu}+\frac{u}{\nu\sqrt{1+z^2}}+O\left(\frac{1}{\nu^2}\right)\right),
 \end{split}\end{equation}
 \begin{equation}\label{eq10_13_9}\begin{split}
uK_{\nu}(\nu z)+\nu z K_{\nu}'(\nu z)=&-\nu \sqrt{1+z^2} K_{\nu}(\nu z)\left(-\frac{u}{\nu\sqrt{1+z^2}}+\frac{1-\frac{v_1}{\nu}+O\left(\frac{1}{\nu^2}\right)}{1-\frac{u_1}{\nu}+O\left(\frac{1}{\nu^2}\right)}\right),\\
 =&-\nu \sqrt{1+z^2}  K_{\nu}(\nu z)\left(1-\frac{v_1-u_1}{\nu}-\frac{u}{\nu\sqrt{1+z^2}}+O\left(\frac{1}{\nu^2}\right)\right),
\end{split} \end{equation}
where
\begin{equation}
u_1=\frac{1}{\sqrt{1+z^2}}\left(\frac{1}{8}-\frac{5 }{24(1+z^2)}\right),   \hspace{1cm}v_1=\frac{1}{\sqrt{1+z^2}}\left(-\frac{3}{8}+\frac{7}{24(1+z^2)}\right).
\end{equation}
We remark that in eq. (16) of \cite{bord08-41-164002}, which corresponds to eq. \eqref{eq10_13_9} here, there is a sign error in $u$ which leads to results different from what we are going to obtain below. From \eqref{eq10_13_12} and \eqref{eq10_13_9}, we have
\begin{equation*}\begin{split}
\frac{uI_{\nu}(\nu z)+\nu z I_{\nu}'(\nu z)}{uK_{\nu}(\nu z)+\nu z K_{\nu}'(\nu z)}=&-\frac{I_{\nu}(\nu z)}{K_{\nu}(\nu z)}\frac{\left(1+\frac{v_1 -u_1 }{\nu}+\frac{u}{\nu\sqrt{1+z^2}}+O\left(\frac{1}{\nu^2}\right)\right)}
{\left(1-\frac{v_1-u_1}{\nu}-\frac{u}{\nu\sqrt{1+z^2}}+O\left(\frac{1}{\nu^2}\right)\right)}\\
=&-\frac{I_{\nu}(\nu z)}{K_{\nu}(\nu z)}\left(1+\frac{2(v_1 -u_1)}{\nu}+\frac{2u}{\nu\sqrt{1+z^2}}+O\left(\frac{1}{\nu^2}\right)\right).\end{split}
\end{equation*}
Thus we obtain  a dependence on the Robin parameter $u$ which is missing in \cite{bord08-41-164002}.

To proceed, we mention that passing from $\text{Y}=\text{D}$ to $\text{Y}=\text{R}$, one has a relative minus sign in front of
$N_{l,l'}^{\text{R}}$. The term $a_{n_i,n_{i+1}}^{(1/2)}$ is not changed, i.e., $a_{n_i,n_{i+1}}^{(1/2),\text{R}}=a_{n_i,n_{i+1}}^{(1/2),\text{D}}$, and the term
$a_{n_i,n_{i+1}}^{(1)}$ is changed by
\begin{equation*}
 a_{n_i,n_{i+1}}^{(1),\text{R}}-a_{n_i,n_{i+1}}^{(1),\text{D}}=\frac{ (\tau^2-1)}{t}+\frac{2u}{t}.
\end{equation*}
Since the change in the sign of $N_{l,l'}$ gives rise to a factor $(-1)^{s+1}$ to $\mathcal{Z}$, it can be compensated by the change in sign of the term $(-1)^{x(s+1)}$ when one passes from $\text{X}=\text{D}$ to $\text{X}=\text{N}$. Therefore we obtain from \eqref{eq10_13_10} that
\begin{equation*}\begin{split}
E^{\text{NR}}-E^{\text{DD}}\sim &-\frac{R}{4\pi d^2}\sum_{s=0}^{\infty}\frac{1}{s+1}\int_0^{\infty}dt \,te^{-2t(s+1)}  \int_0^1\frac{d\tau\sqrt{\tau}}{\sqrt{1-\tau^2}}\int_{-\infty}^{\infty}\frac{d\mu}{\sqrt{\pi}}e^{-\mu^2(s+1)/\tau}
\left(\prod_{j=1}^s\int_{-\infty}^{\infty}\frac{dn_j}{\sqrt{\pi}}\right) e^{-\eta_1}\\&\hspace{4cm}\times\left(\vep\sum_{i=0}^s\left[\frac{ (\tau^2-1)}{t}+\frac{2u}{t}\right]\right).\end{split}\end{equation*}
This integral is easy to compute and we find that
\begin{equation*}
E^{\text{NR}}-E^{\text{DD}}\sim -\frac{R}{16\pi d^2}\sum_{s=0}^{\infty}\frac{\vep}{(s+1)^2}\frac{2(6u-1)}{3}.
\end{equation*}Therefore,
\begin{equation*}\begin{split}
E^{\text{NR}}= &-\frac{R}{16\pi d^2  }\sum_{s=0}^{\infty}\left(\frac{1}{(s+1)^4}\left(1+\frac{\vep}{3} \right)+\frac{\vep}{(s+1)^2}\frac{2(6u-1)}{3}+\ldots\right)
\\=&-\frac{\pi^3 R}{1440 d^2}\left(1+\left[\frac{1}{3}+\frac{10(6u-1)}{\pi^2}\right]\vep+\ldots\right),
\end{split}\end{equation*}where we have used
$$\sum_{s=0}^{\infty}\frac{1}{(s+1)^2}=\zeta(2)=\frac{\pi^2}{6}.$$
Similarly, we have
\begin{equation*}
E^{\text{DR}}-E^{\text{ND}}\sim -\frac{R}{16\pi d^2}\sum_{s=0}^{\infty}\frac{\vep(-1)^{s+1}}{(s+1)^2}\frac{2(6u-1)}{3}.
\end{equation*}This implies that
\begin{equation*}\begin{split}
E^{\text{DR}}= &-\frac{R}{16\pi d^2  }\sum_{s=0}^{\infty}\left(\frac{(-1)^{s+1}}{(s+1)^4}\left(1+\frac{\vep}{3} \right)+\frac{\vep(-1)^{s+1}}{(s+1)^2}\frac{2(6u-1)}{3}+\ldots\right)\\
=&\frac{7\pi^3 R}{11520 d^2}\left(1+\left[\frac{1}{3}+\frac{40(6u-1)}{7\pi^2}\right]\vep+\ldots\right),
\end{split}\end{equation*}where   the formula
$$\sum_{s=0}^{\infty}\frac{(-1)^{s+1}}{(s+1)^2}=-\frac{1}{2}\frac{\pi^2}{6}$$ has been applied.

The results in this section can be summarized as
\begin{equation}\label{sc1}
\begin{split}
\frac{E^{\text{DD}}}{E_{\text{PFA}}^{\text{DD}}}=&1+\frac{\vep}{3}+\ldots
%\simeq 1+0.33\vep+\ldots
,\\
\frac{E^{\text{ND}}}{E_{\text{PFA}}^{\text{ND}}}=&1+\frac{\vep}{3}+\ldots
%\simeq 1+0.33\vep+\ldots
,\\
\frac{E^{\text{DR}}}{E_{\text{PFA}}^{\text{DR}}}=&1+\left(\frac{1}{3}+\frac{80(3\alpha-2)}{7\pi^2}\right)\vep+\ldots
%\simeq 1+(3.47\alpha-1.98)\vep+\ldots
,\\
\frac{E^{\text{NR}}}{E_{\text{PFA}}^{\text{NR}}}=&1+\left(\frac{1}{3}+\frac{20(3\alpha-2)}{ \pi^2}\right)\vep+\ldots
%\simeq 1+(6.08\alpha-3.72)\vep+\ldots
.
\end{split}
\end{equation}
Here, $E_{\text{PFA}}^{\text{XY}}$ is the leading term which coincides with the proximity force approximation,
\begin{equation*}
E_{\text{PFA}}^{\text{DD}}=E_{\text{PFA}}^{\text{NR}}=-\frac{\pi^3 R}{1440 d^2},\hspace{1cm}
E_{\text{PFA}}^{\text{ND}}=E_{\text{PFA}}^{\text{DR}}= \frac{7\pi^3 R}{11520 d^2}.
\end{equation*}
Setting $\alpha$ equal to zero,  which corresponds to the Neumann case, we find that
\begin{equation}\label{sc2}
\begin{split}
\frac{E^{\text{DN}}}{E_{\text{PFA}}^{\text{DN}}}=&1+\left(\frac{1}{3}-\frac{160}{7\pi^2}\right)\vep+\ldots\simeq 1-1.98\,\vep+\ldots,\\
\frac{E^{\text{NN}}}{E_{\text{PFA}}^{\text{NN}}}=&1+\left(\frac{1}{3}-\frac{40}{ \pi^2}\right)\vep+\ldots\simeq 1-3.72\,\vep+\ldots.
\end{split}
\end{equation}
%The results for the DD, NN, DN and ND cases have been obtained recently in \cite{bimo11-1110.1082} using derivative expansion. Our results here agree completely with those obtained in that paper.
We mention that the corrections involving Neumann boundary conditions on the sphere are still quite large.

\section{The electromagnetic field}
For electromagnetic field, if the sphere and the plane are both perfectly conducting, the Casimir energy is given by \cite{emig08-04007,bord10-81-065011}:
\begin{equation*}
E^{\text{EM}}=\frac{1}{2\pi}\int_0^{\infty}d\xi\,\text{Tr}\ln \left(1-\mathbb{N}(\xi)\right).
\end{equation*}The trace Tr is
\begin{equation*}
\text{Tr} =\sum_{m=-\infty}^{\infty}\sum_{l=\max\{1,|m|\}}^{\infty}\text{tr},
\end{equation*}where the trace tr on the right hand side is the trace over  $2\times 2$-matrices -- the components of $\mathbb{N}$ given by
\begin{equation*}\begin{split}
\mathbb{N}_{l,l'}=&\sqrt{\frac{\pi}{4\xi L}}\sum_{l^{\prime\prime}=|l-l'|}^{l+l'}K_{l^{\prime\prime}+1/2}(2\xi L)H_{ll'}^{l^{\prime\prime}}
\begin{pmatrix}\Lambda_{l,l'}^{l^{\prime\prime}} & \tilde{\Lambda}_{l,l'}\\ \tilde{\Lambda}_{l,l'} &\Lambda_{l,l'}^{l^{\prime\prime}}\end{pmatrix}
\begin{pmatrix} d_l^{\text{TE}}(\xi R) & 0\\0 & -d_l^{\text{TM}}(\xi R)\end{pmatrix} ,
\end{split}\end{equation*}
with
\begin{equation*}
\Lambda_{l,l'}^{l^{\prime\prime}}=\frac{1}{2}\frac{l^{\prime\prime}(l^{\prime\prime}+1)-l(l+1)-l'(l'+1)}{\sqrt{l(l+1)l'(l'+1)}},\hspace{1cm}
 \tilde{\Lambda}_{l,l'}=\frac{2m \xi L}{\sqrt{l(l+1)l'(l'+1)}}
\end{equation*}and
\begin{equation*}
d_l^{\text{TE}}( \xi R)=\frac{I_{l+1/2}( \xi R)}{K_{l+1/2}( \xi R)},\hspace{1cm} d_l^{\text{TM}}( \xi R)=\frac{\frac{1}{2}I_{l+1/2}( \xi R)+ \xi RI_{l+1/2}'( \xi R)}
{\frac{1}{2}K_{l+1/2}( \xi R)+ \xi RK_{l+1/2}'( \xi R)}.
\end{equation*}
Notice that $d_l^{\text{TE}}$ and $d_l^{\text{TM}}$ correspond, respectively, to $d_l^{\text{D}}$ and $d_l^{\text{R}}$ with $u=1/2$ in the scalar case.

Proceeding as in the previous section, we find that the Casimir energy can be written in the same form as \eqref{eq10_13_4}, with $\mathcal{Z}$ now given by
\begin{equation}\label{eq10_14_1}
\mathcal{Z} =\text{tr}\prod_{i=0}^s\left(\sqrt{\frac{4\pi t}{\vep}}\mathbb{N}_{l+\tilde{l_i},l+\tilde{l}_{i+1}}\right).
\end{equation}
To expand $\mathcal{Z}$ up to terms of order $\vep$, we only need, in addition to the scalar case, to expand the diagonal term $\Lambda_{l,l'}^{l^{\prime\prime}}$ up to terms of order $\vep$, and the off-diagonal term $\tilde{\Lambda}_{l,l'}$ up to terms of order $\sqrt{\vep}$. Using the substitutions \eqref{eq10_14_2}, we have
\begin{equation*}\begin{split}
 \Lambda_{l,l'}^{l^{\prime\prime}}=&1+\lambda_{n,n'}(\nu)\vep+\ldots,\hspace{1cm}\lambda_{n,n'}(\nu):= -\frac{1}{t\tau}-\frac{4\nu}{t\tau},\\
 \tilde{\Lambda}_{l,l'}=&\tilde{\lambda}_{n,n'}\sqrt{\vep}+\ldots,\hspace{2cm}\tilde{\lambda}_{n,n'}:= \frac{2\mu\sqrt{1-\tau^2}}{\sqrt{t}\tau^{3/2}}.
\end{split}\end{equation*}
Comparing to the scalar case, it is easy to see that the asymptotic expansion for $\mathbb{N}_{l,l'}$ is given by
\begin{equation*}
\mathbb{N}_{l,l'}\sim \sqrt{\frac{\vep}{4\pi t}}e^{-2t-(n-n')^2-\mu^2/\tau}
\left\{\begin{pmatrix}1 & 0\\ 0 & 1\end{pmatrix}+\sqrt{\vep}\begin{pmatrix} a_{n,n'}^{(1/2),\text{TE}} & \tilde{\lambda}_{n,n'} \\\tilde{\lambda}_{n,n'} & a_{n,n'}^{(1/2),\text{TM}}\end{pmatrix}+\vep\begin{pmatrix}a_{n,n'}^{(1),\text{TE}}+\hat{\lambda}_{n,n'} & *\\ * &a_{n,n'}^{(1),\text{TM}}+\hat{\lambda}_{n,n'}
\end{pmatrix}+\ldots\right\}
\end{equation*}where
\begin{equation*}
a_{n,n'}^{(i),\text{TE}}=a_{n,n'}^{(i),\text{D}},\hspace{1cm}a_{n,n'}^{(i),\text{TM}}=a_{n,n'}^{(i),\text{R}}\bigr|_{u=1/2}
\end{equation*}for $i=1/2$ or $1$; and
\begin{equation}\label{eq10_14_5}
\hat{\lambda}_{n,n'}=\sqrt{\frac{2\tau}{ 1+\tau}}e^{\mu^2/\tau}\int_{-\infty}^{\infty}\frac{d\eta}{\sqrt{\pi}}e^{-\eta^2+2i\eta\sqrt{2}\mu +\mu^2 }\sum_{\nu=0}^{\infty}\frac{\eta^{2\nu}}{\nu!}
\left(\frac{1-\tau}{1+\tau}\right)^{\nu}\lambda_{n,n'}(\nu).
\end{equation}
The asterisks denote terms contributing to higher orders.

Substituting into \eqref{eq10_14_1} and taking the trace, we find that
\begin{equation*}
\mathcal{Z}=2+\sqrt{\vep}\left(\sum_{i=0}^s\left[a_{n_i,n_{i+1}}^{(1/2),\text{TE}}+a_{n_i,n_{i+1}}^{(1/2),\text{TM}}\right]\right)+\vep\left(a^{\text{TE}}+a^{\text{TM}}+2\sum_{i=0}^s\hat{\lambda}_{n_i,n_{i+1}}+2\sum_{0\leq i<j\leq s}\tilde{\lambda}_{n_i,n_{i+1}}\tilde{\lambda}_{n_j,n_{j+1}}\right)+\ldots.
\end{equation*}
Here $a^{\text{TE}}=a^{\text{D}}$ (eq. \eqref{eq10_14_4}) and $a^{\text{TM}}$ is defined in the similar way. The term of order $\sqrt{\vep}$  will drop out after integration with respect to $n_i$. Substituting into \eqref{eq10_13_4} and compare to \eqref{eq10_13_10}, one finds that the first two leading terms of the electromagnetic Casimir energy can be written as
\begin{equation*}
E^{\text{EM}}=
\left\{
\begin{aligned}\text{first two leading}\\\text{terms of }\, E^{\text{DD}\quad}\end{aligned}
\right\}+
\left\{
\begin{aligned}\text{first two leading}\quad\\\text{terms of }\,E^{\text{NR} }\bigr|_{u=1/2}\end{aligned}
\right\}+\Delta E,
\end{equation*}
where
\begin{equation}\label{eq10_14_6}\begin{split}
\Delta E=&-\frac{R}{4\pi d^2}\sum_{s=0}^{\infty}\frac{1}{s+1}\int_0^{\infty} dt\,te^{-2t(s+1)}  \int_0^1\frac{d\tau\sqrt{\tau}}{\sqrt{1-\tau^2}}\int_{-\infty}^{\infty}\frac{d\mu}{\sqrt{\pi}}e^{-\mu^2(s+1)/\tau}
\left(\prod_{j=1}^s\int_{-\infty}^{\infty}\frac{dn_j}{\sqrt{\pi}}\right)\\&
\hspace{3cm}\times e^{-\eta_1}\vep \left( 2\sum_{i=0}^s\hat{\lambda}_{n_i,n_{i+1}}+2\sum_{0\leq i<j\leq s}\tilde{\lambda}_{n_i,n_{i+1}}\tilde{\lambda}_{n_j,n_{j+1}}\right).\end{split}
\end{equation}

Up to this point, it seems that we haven't done anything much different from the one presented in the paper \cite{bord10-81-065011}. The major difference here is that we only keep the terms in $\Lambda_{l,l'}^{l^{\prime\prime}}$ and $\tilde{\Lambda}_{l,l'}$ that will contribute terms up to order $\vep$ for $\mathcal{Z}$. In the following, we are going to see that this simplifies the computations.

Using
\begin{equation*}
\sum_{\nu=0}^{\infty}\frac{\nu \eta^{2\nu}}{\nu!}
\left(\frac{1-\tau}{1+\tau}\right)^{\nu}=\frac{1-\tau}{1+\tau}\eta^2\exp\left(\frac{1-\tau}{1+\tau}\eta^2\right),
\end{equation*}it is straightforward to compute $\hat{\lambda}_{n,n'}$ given by \eqref{eq10_14_5}. We find that
\begin{equation*}
\hat{\lambda}_{n,n'}=\frac{2(1-\tau^2)}{t\tau^3}\mu^2-\frac{1}{t\tau^2}.
\end{equation*}Then
\begin{equation}\label{a1}\begin{split}
&2\sum_{i=0}^s\hat{\lambda}_{n_i,n_{i+1}}+2\sum_{0\leq i<j\leq s}\tilde{\lambda}_{n_i,n_{i+1}}\tilde{\lambda}_{n_j,n_{j+1}} \\=&2(s+1)\left(\frac{2(1-\tau^2)}{t\tau^3}\mu^2-\frac{1}{t\tau^2}\right)
+s(s+1)\left(\frac{2\mu\sqrt{1-\tau^2}}{\sqrt{t}\tau^{3/2}}\right)^2\\
=&\frac{4(s+1)^2(1-\tau^2)}{t\tau^3}\mu^2-\frac{2(s+1)}{t\tau^2}.
\end{split}\end{equation}
Substituting  into \eqref{eq10_14_6} and carrying out the $\mu$-integration, it is seen that the singularity for $\tau\to0$ cancels between both contributions in the last line in \Ref{a1}. This is equivalent to the compensation of the logarithms observed in \cite{bord10-81-065011} (see remark after eq.(135)), however now the remaining contributions come out different.
We find that
\begin{equation*}
\Delta E=\frac{R}{4\pi d^2}\sum_{s=0}^{\infty}\frac{\vep}{(s+1)^2}=\frac{R}{4\pi d^2}\frac{\pi^2}{6}\vep.
\end{equation*}Using the results for $E^{\text{DD}}$ and $E^{\text{NR}}\bigr|_{u=1/2}$ from the previous section, we finally obtain the first two leading terms for the electromagnetic Casimir energy as
\begin{equation}\label{elm}\begin{split}
E^{\text{EM}}=&-\frac{\pi^3 R}{1440 d^2}\left(1+\frac{\vep}{3} \right)-\frac{\pi^3 R}{1440 d^2}\left(1+\left[\frac{1}{3}+\frac{20}{\pi^2}\right] \vep\right)
+\frac{R}{4\pi d^2}\frac{\pi^2}{6}\vep+\ldots\\
=&-\frac{\pi^3 R}{720 d^2}\left(1+\left[\frac{1}{3}-\frac{20}{\pi^2}\right]\vep+\ldots\right)\\
\simeq & 1-1.69\vep+\ldots.
\end{split}
\end{equation}This result agrees with that obtained recently in \cite{bimo11-1110.1082} using derivative expansion. As has been observed in \cite{bimo11-1110.1082},   the first two leading terms of the EM Casimir energy for perfect conductors  turns up to be the sum of the first two leading terms for the DD and the NN scalar Casimir energies.

\section{Conclusions}
The results from the calculation of the first correction beyond PFA are summarized in Eqs. \Ref{sc1}, \Ref{sc2} for the scalar field and in \Ref{elm} for the electromagnetic field. These formulas substitute the final formulas  Eq. (3) in \cite{bord08-41-164002} and Eqs. (43) and (136) in \cite{bord10-81-065011}. It can be seen that the resulting numbers  change to some extent and that the logarithmic contributions disappeared.

Very recently, in \cite{fosc11-1109.2123} and in \cite{bimo11-1110.1082}, another method for obtaining analytic corrections beyond PFA was found. It consists of an expansion of $E$, \Ref{E0}, which is perturbative in the gradients of the height profiles of the interacting surfaces. Intuition predicts that neglecting the derivatives, a resummation of the perturbative expansion in the height profiles, should reproduce PFA, while the inclusion of gradients should give the first corrections beyond PFA \cite{fosc11-1109.2123}. Indeed, in \cite{fosc11-1109.2123}, the analytic result for Dirichlet boundary conditions was obtained confirming the first line in Eq. \Ref{sc1}. All other cases were computed in \cite{bimo11-1110.1082} and coincide with Eqs. \Ref{sc1}, \Ref{sc2} and \Ref{elm}.
In \cite{bimo11-1110.1082}, a Pad\'e resummation of the asymptotic large distance expansion, matched with the first PFA correction, is performed and
yields excellent agreement with the numerical results in \cite{bimo11-1110.1082} for DD, NN and EM boundary conditions. In addition, fits to the numerical data in \cite{bimo11-1110.1082},
that take into account logarithmic corrections to PFA at second order (in distinction to \cite{emig08-04007}), yield results that are consistent with the results given here for the three types of boundary conditions.

\begin{acknowledgments}\noindent
One of the authors (MB) acknowledges a helpful discussion with T. Emig.
\\ LPT is supported by the Ministry of Higher Education of Malaysia  under the FRGS grant FRGS/2/2010/SG/UNIM/02/2.
\end{acknowledgments}
%\begin{thebibliography}{10}
%\bibitem{1} M. Bordag and V. Nikolaev,  J. Phys. A \textbf{41}, 164002 (2008).
%\bibitem{2} M Bordag and V Nikolaev,   Phys. Rev. D \textbf{81}, 065011 (2010).
%\bibitem{3} A. Bulgac, P. Magierski  and A. Wirzba, Phys. Rev. D \textbf{73}, 025007 (2006).
%\bibitem{4} M. Bordag,   Phys. Rev. D \textbf{73}, 125018 (2006).
%\bibitem{5} T. Emig, J. Stat. Mech. \textbf{2008}, P04007 (2008).
%\bibitem{6} G. Bimonte, T. Emig, R. L. Jaffe and M. Kardar, arXiv:1110.1082.
% \end{thebibliography}

\bibliographystyle{unsrt}%\bibliography{C:/Users/bordag/WORK/Literatur/bib/papers,C:/Users/bordag/WORK/Literatur/Bordag}

\end{document}